\documentclass[11pt]{article}
\usepackage{amsmath,amssymb}
\usepackage{epsfig,graphicx}

\long\def\inst#1{\par\nobreak\kern 4pt\nobreak
    {\it #1}\par\vskip 10pt plus 3pt minus 3pt}

\RequirePackage{xspace}
\def\babar{{\em B}{\footnotesize\em A}{\em B}{\footnotesize\em AR}}
\def\CP                {\ensuremath{C\!P}\xspace}
\def\CM                {\ensuremath{C\!M}\xspace}
\def\FB                {\ensuremath{F\!B}\xspace}

\def\Kbar  {\kern 0.2em\overline{\kern -0.2em K}{}\xspace}

\def\Kz    {\ensuremath{K^0}\xspace}
\def\Kzb   {\ensuremath{\Kbar^0}\xspace}
\def\KzKzb {\ensuremath{\Kz \kern -0.2em - \kern -0.2em \Kzb}\xspace}
\def\KS    {\ensuremath{K^0_{\scriptscriptstyle S}}\xspace}
\def\KL    {\ensuremath{K^0_{\scriptscriptstyle L}}\xspace}
\def\KSKL {\ensuremath{\KS \kern -0.2em - \kern -0.2em \KL}\xspace}

\def\epem  {\ensuremath{e^+\!e^-}\xspace}
\def\Dz      {\ensuremath{D^0}\xspace}
\def\Dbar    {\kern 0.18em\overline{\kern -0.18em D}{}\xspace}
\def\Kz      {\ensuremath{K^0}\xspace}
\def\Kzb     {\ensuremath{\Kbar^0}\xspace}
\def\Kpm   {\ensuremath{K^\pm}\xspace}
\def\pipm  {\ensuremath{\pi^\pm}\xspace}
\def\pip     {\ensuremath{\pi^+}\xspace}
\def\calA   {\ensuremath{\cal A}\xspace}
\def\calM   {\ensuremath{\cal M}\xspace}

\newcommand{\Dtoksk}{\ensuremath{D^{\pm}\to\KS\Kpm}\xspace}
\newcommand{\Dstoksk}{\ensuremath{D_s^{\pm}\to\KS\Kpm}\xspace}
\newcommand{\Dstokspi}{\ensuremath{D_s^{\pm}\to\KS\pipm}\xspace}
\newcommand{\kspipm}{\ensuremath{\KS\pipm}\xspace}
\newcommand{\etal}{{\em et al.}}
\def\babar{{\em B}{\footnotesize\em A}{\em B}{\footnotesize\em AR}}

\def\Dzb     {\ensuremath{\Dbar^0}\xspace}
\def\Dp      {\ensuremath{D^+}\xspace}
\def\Dm      {\ensuremath{D^-}\xspace}
\def\Dpm      {\ensuremath{D^{\pm}}\xspace}
\def\Dstp      {\ensuremath{D^{*+}}\xspace}
\def\Dspm      {\ensuremath{D^{\pm}_{s}}\xspace}
\def\D(s)pm      {\ensuremath{D^{\pm}_{(s)}}\xspace}
\def\CKM                {\ensuremath{C\!K\!M}\xspace}
\def\CPV                {\ensuremath{C\!P\!V}\xspace}
\def\SM                {\ensuremath{S\!M}\xspace}
\def\NP                {\ensuremath{N\!P}\xspace}
\def\ra                 {\ensuremath{\rightarrow}\xspace}
\def\CP                {\ensuremath{C\!P}\xspace}
\def\DP                {\ensuremath{D\!P}\xspace}

\newcommand{\DpmtoKKpipm}{\ensuremath{\Dpm \ra K^+K^- \pipm}\xspace}
\newcommand{\DptoKKpip}{\ensuremath{\Dp \ra K^+K^- \pip}\xspace}

\newcommand{\jprlBase}       {Phys.\ Rev.\ Lett.\xspace}
\newcommand{\jprl}      [1]  {\jprlBase\ {\bf #1}}
\newcommand{\jprBase}        {Phys.\ Rev.\xspace}
\newcommand{\jprd}      [1]  {\jprBase\ D~{\bf #1}}
\newcommand{\jplBase}        {Phys.\ Lett.\xspace}
\newcommand{\plb}       [1]  {\jplBase\ B~{\bf #1}}
\newcommand{\nimBaseA}       {Nucl.\ Instr.\ Meth.\xspace}
\newcommand{\nima}      [1]  {\nimBaseA~A~{\bf #1}}
\newcommand{\jpBase}       {J.\ Phys.\xspace}
\newcommand{\jp}      [1]  {\jpBase\ G ~{\bf #1}}

\def\ACP  {\ensuremath{{\cal A}_{\CP}}\xspace}
\def\AFB  {\ensuremath{{\cal A}_{\FB}}\xspace}
\def\AEP  {\ensuremath{{\cal A}_{\epsilon}}\xspace}

\begin{document}
\thispagestyle{empty}

\begin{flushright}
October 5, 2012 \\
\end{flushright}

\par\vskip 4cm

\begin{center}
\Large \bf Searches for New Physics in \CP Violation from \babar 
\end{center}
\bigskip

\begin{center}
\large 
F. Palombo\\
Universit\`a di Milano, Dipartimento di Fisica and INFN, I-20133 Milano, Italy \\
(for the \babar\ Collaboration)
\end{center}
\bigskip \bigskip

\begin{center}\large \bf Abstract
\end{center}
Results of recent searches  for new physics in  \CP violation  in  charm decays from the \babar\ experiment are presented. 
These results  include a measurement of   \Dz\ - \Dzb  mixing and  searches  for  
\CP violation in two-body \Dz\  decays, a  search for \CP violation in the charm  decays  \Dtoksk and \Dstoksk,   \kspipm , and a search for  direct \CP violation  in  the singly-Cabibbo suppressed $D^{\pm} \ra  K^+ K^- \pi^{\pm}$  decays. These studies are based on the final  dataset  collected  by \babar\   at the PEP-II   B factory at SLAC in the period 1999-2008. No evidence of  \CP 
violation is found in  these charm decays.  The measured mixing parameter $y_{\CP} = [0.72 \pm  0.18 (stat) \pm 0.12 (syst)]$\% excludes the  no-mixing null hypothesis  with  a significance of  3.3$\sigma$.

\vfill
\begin{center}
Presented at QUARKS-2012,  the 17th International Seminar on High Energy Physics, \\
Yaroslavl, Russia, 4-10 June, 2012.
\end{center}

\newpage
\pagenumbering{arabic}
\setcounter{page}{1}

\section{Introduction}

The  Cabibbo-Kobayashi-Maskawa (\CKM)  paradigm of  \CP violation (\CPV) in the Standard Model (\SM) has been tested by  the \babar\  and Belle experiments with high precision in many overconstrained measurements.  Nevertheless  \SM  leaves many unanswered questions.  \CPV  is one of
the three Sakharov necessary conditions to generate the asymmetry between matter-antimatter (baryogenesis)  observed in the Universe.
The measured \CKM\ weak phase  is unable to provide enough \CPV\ to explain the observed baryon asymmetry. 
New \CPV sources are needed from New Physics (\NP)  beyond the \SM.  

 At the B-factories important areas of search for  \NP\   are  processes which are expected at low level in \SM and which could be enhanced by \NP.   In these \NP searches  at  low energies, charm physics plays currently  an important and increasing role. The \babar, Belle and CDF measurements  of flavor mixing in the neutral D meson system~\cite{Dmixing} show evidence of  \Dz\ - \Dzb  mixing at 1 \% level. 
   These results  are  in agreement with \SM predictions~\cite{Buccella,Bianco,Gross} and sets  constraints on possible   contributions  from  many \NP models~\cite{Golo}.

Recently the LHCb collaboration has reported  a first  evidence of  \CP violation  in \Dz  decays to $K^+ K^-$ and $\pi^+ \pi^-$~\cite{LHCb}.  This evidence has been confirmed by the CDF Collaboration~\cite{CDF}. Given the \SM expectation that \CPV in charm sector  should be at the level  of $10^{-3}$ (or lower)~\cite{Buccella,Gross}, this \CPV evidence was rather unexpected at  the present experimental
sensitivity. Marginally compatible with the  \SM expectations,  this  \CPV  may be a manifestation of   \NP or of significant enhancements of penguin diagrams in charm 
decays~\cite{Gross2,Isidori,Franco,Cheng,Gronau}.  

In this talk I  present recent \babar\ measurements concerning mixing and \CP violation in charm sector.    All new results presented here are preliminary.

 \section {\Dz - \Dzb Mixing and CP Violation in Two-Body \Dz  decays}
 Mixing in charm sector is unique because   it involves virtual down type quarks. It arises from both short range and long  range contributions. The short range contribution is expected to be very small because of \CKM and GIM suppressions. The dominant long range contribution is non perturbative and hard to evaluate. This implies large theoretical 
 uncertainties in the \SM calculations of the mixing parameters $x$ and $y$~\cite{Bianco,Xing,Burd}.   \SM expectations values for $x$ and $y$ are $\leq 10^{-3}$ but higher values  are predicted in some \NP  models~\cite{Petrov,Gersa}.

 In  a recent \babar\  analysis~\cite{Casarosa}  charm mixing and \CP violation are  measured  using the  ratio of lifetimes obtained in the  \Dz   decays to the two-body final states $K^{\mp} \pi^{\pm}$, $K^- K^+$, and  $\pi^- \pi^+$. The analysis is based on an integrated luminosity of 468 $fb^{-1}$ collected by the \babar\   detector~\cite{babar}. Five  different signal channels are  considered~\cite{CC}: three  flavor tagged channels  $\Dstp \ra \Dz \pi^+_s$  with $\Dz  \ra K^+ K^-$,  $\Dz  \ra \pi^+ \pi^-$, and  $\Dz \ra K^- \pi^+,  K^+ \pi^-$  and 
two flavor untagged channels $\Dz \ra K^+ K^-$ and $\Dz  \ra K^- \pi^+$, $K^+ \pi^-$, where $\pi^+_s$ is a slow pion track used in the tagging algorithm. 
 
 The  experimental observables $y_{CP}$~\cite{Liu} , sensitive to mixing, and $\Delta Y$, sensitive to \CPV\ ,  are measured. These observables are defined as:
 
 \begin{eqnarray}
 \label{eq:Mix}
 y_{CP} \equiv \frac{ \Gamma^+ + \bar{\Gamma}^+ }{ \Gamma }  -1  \qquad  and \qquad  \Delta Y \equiv \frac{ \Gamma^+ -  \bar{\Gamma}^+ }{ 2 \Gamma }\, ,
 \end{eqnarray}
 where $\Gamma^+$ ($\bar{\Gamma}^+$) is the average width of the \Dz (\Dzb) when reconstructed in the \CP-even eigenstates ($K^+ K^-$, $\pi^+ \pi^-$).  $\Gamma$ is the average \Dz width
 describing the decays to the CP-mixed final states $K^{\mp}\pi^{\pm}$~\cite{Previous}.
 
 Neglecting  contribution of direct \CP violation estimated at a level below   our sensitivity~\cite{Gersak} and taking into account that the \CP violating  weak phase $\phi$ in \SM to a good approximation does not depend on the final states~\cite{Soko}, the observables $y_{CP}$ and $\Delta Y$ in terms of the mixing parameters $x$ and $y$ can be written as:
\begin{eqnarray}
\label{eq:array}
 y_{CP} &=& y \cos \phi + \frac{{\cal A}_M}{2} x \sin \phi  \nonumber\\
 \Delta Y &=& -x \sin \phi + \frac{{\cal A}_M}{2} y \cos \phi \,,
 \end{eqnarray}
 where    ${\cal A}_M =  ( |q/p|^2 - |p/q|^2) / (  |q/p|^2 + |p/q|^2 )  $  measures  the \CP asimmetry in mixing.  The complex parameters $p$ and $q$ relate the mass eigenstates of neutral mesons , $|D_{1,2}>$,  to the  flavor eigenstates,  $|\Dz>$  and $|\Dzb>$,  through the relation  $|D_{1,2}> =  p   |\Dz>  \pm q |\Dzb>$.
 
\begin{figure}[htb]
\begin{center}
\hspace{-9cm}
\includegraphics[scale=0.32] {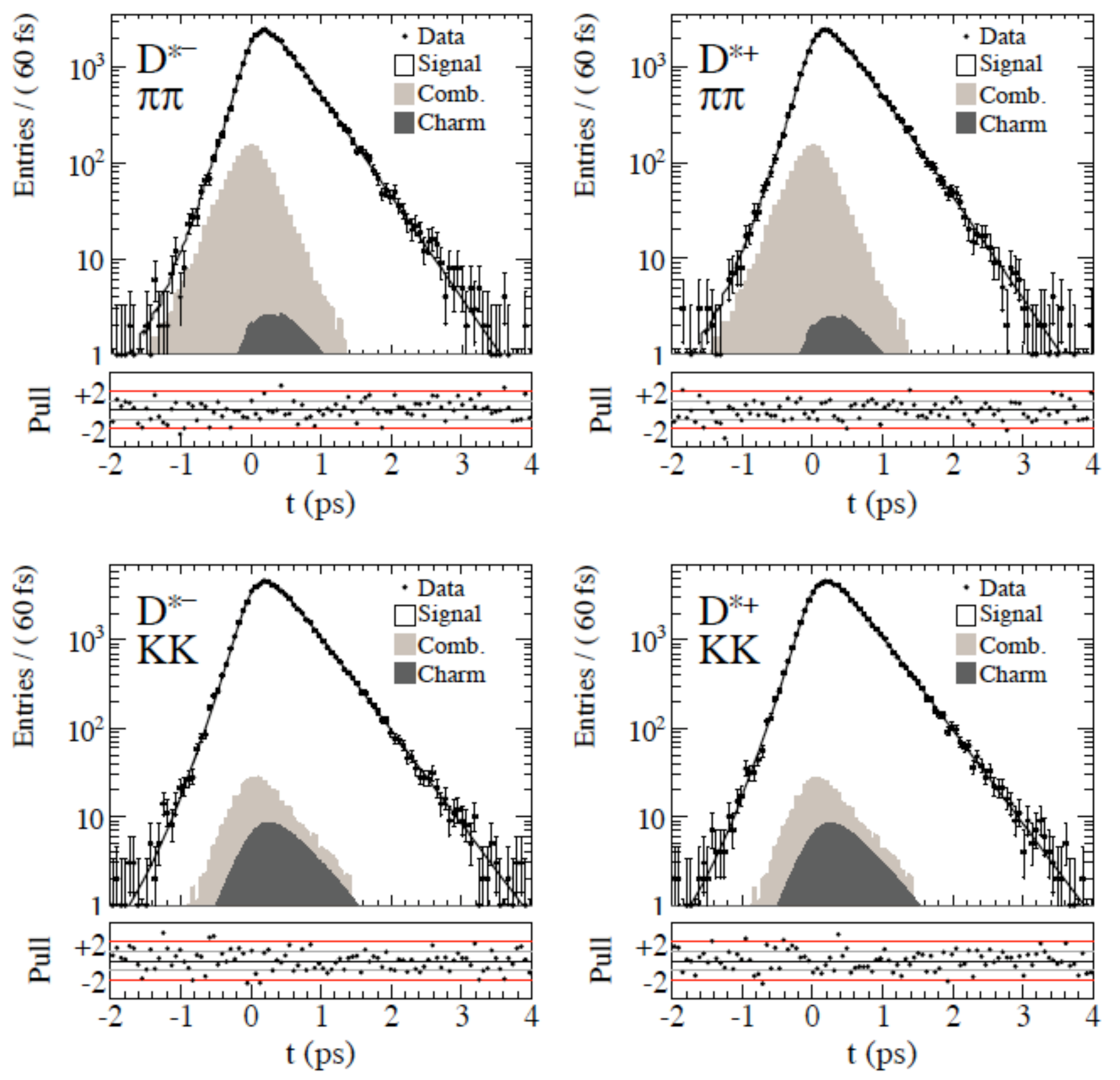} \\
\vspace{-7.2cm}
\hspace {7cm}
\includegraphics[scale=0.32] {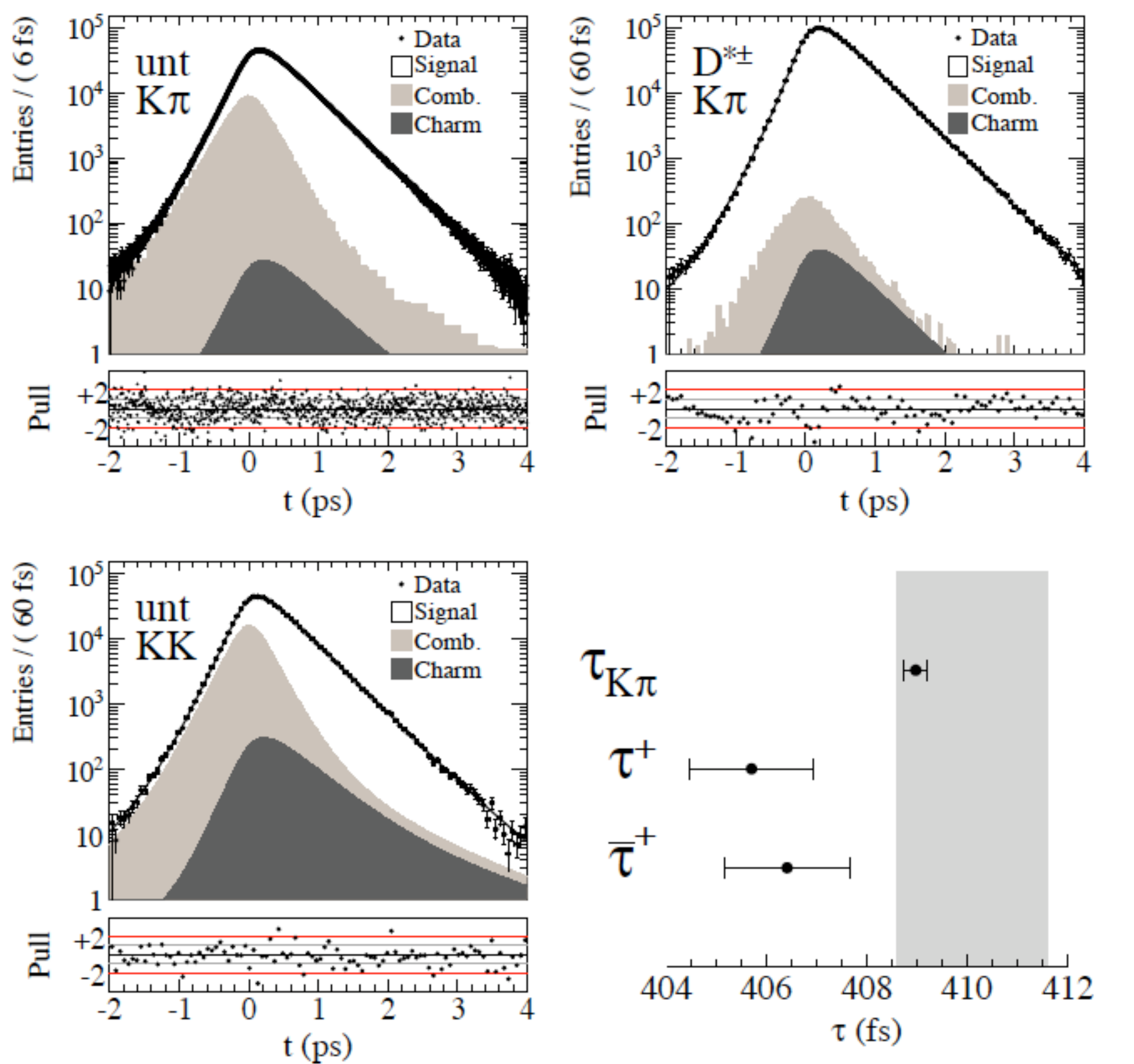} \\
\caption{ Proper time fit projections  with the fit results overlaid. The combinatorial background distribution (Comb) is stacked on the top  of the misreconstructed-charm background distribution (Charm). Under each plot are shown the normalized Poisson pulls; "unt" refers to the untagged dataset.  The gray band is the PDG \Dz lifetime $\pm \sigma$\cite{PDG}. }
\label{fig:Tdistr}
\end{center}
\end{figure}

A simultaneous  extended unbinned Maximum Likelihood (ML) fit to the two-dimensional distribution of the proper time and proper time error in 
tagged and untagged modes  is performed:  the average \Dz lifetime $\tau$ is extracted from $K^{\mp} \pi^{\pm}$ final states and the effective lifetime $\tau^+$ ($\bar{\tau}^+$)  is extracted from \Dz (\Dzb) decays to the final states $K^-K^+$ and $\pi^-\pi^+$.  Main sources of background are misreconstructed charm events and the combinatorial background candidates  consisting of random tracks.
Using the reciprocals of the   three measured lifetimes in Eq.~\ref{eq:Mix}  we obtain:

$$
y_{\CP} = (0.72 \pm 0.18 \pm 0.12) \% \qquad \quad and \qquad \quad \Delta Y = (0.09 \pm 0.26 \pm 0.09) \% \,,
$$
where the first uncertainty  is statistical and the second systematic. Projections of lifetime fit are  shown in Fig.~\ref{fig:Tdistr}.

These results exclude  no-mixing hypothesis  at 3.3$\sigma$ significance and show no evidence of  \CPV.  The $y_{CP}$ value  is consistent  with the mixing parameter y  measured in the decays $\Dz \ra \KS h^+ h^-$ (where $h = K, \pi$)~\cite{Asner}  as expected in absence of \CPV.
This $y_{CP}$ measurement is the most precise single  measurement up to date.  These results are in agreement with \SM predictions.

 \section {\CP Violation in \Dtoksk and   \Dstoksk, \kspipm} 
 
 The channels $\Dpm \ra  \KS K^{\pm}$ can proceed through Cabibbo-Favored (CF) and Doubly-Cabibbo-Suppressed (DCS) transitions. The CF transition  is largely
 dominating and the \SM expectation for direct  \CP is negligible.  The channels  $ \Dspm \ra \KS K^{\pm}\, , \KS \pi^{\pm}$ can proceed through two Singly-Cabibbo-Suppressed (SCS) transitions, both of comparable amplitudes.  The relative 
  phase between the  two decay amplitudes can generate  interference effects and induce direct \CPV.

 In these channels with a \Kz (or \Kzb) in the final state a time-integrated \CPV\  of  $\approx  (\pm 0.332 \pm 0.006)\, $\%   is induced by the  $\Kz \Kzb\  $  mixing~\cite{PDG}. The sign of this asymmetry is positive (negative) in presence in the final state of a \Kz\  (\Kzb). The exact value of this \CPV  asymmetry contribution depends on the requirements on the reconstructed $\KS \ra \pi^+ \pi^-$ decays and the decay kinematics~\cite{Nir}.   
 
Previous results of searches for direct  \CP violation in these  decay modes  by CLEO-c~\cite{Cleo} and Belle~\cite{Belle} Collaborations  are all in agreement with \SM expectations.   

Direct \CP asymmetry in these charm decay modes has been recently  searched for  by \babar~\cite{Cenci} using a dataset of $469 fb^{-1}$. The following  direct \CP-violating parameter  \ACP\  is measured for each decay channel:

\begin{equation}
\ACP =  \frac{\Gamma\large(D^+_{(s)} \ra  \KS (\pi^+, K^+) \large)  - \Gamma \large(D^-_{}(s) \ra  \KS (\pi^+, K^+)   )}  {\Gamma\large(D^+_{(s)} \ra  \KS (\pi^+, K^+) \large)  + \Gamma \large(D^+_{(s)} \ra  \KS (\pi^+, K^+)   )} \,,
 \end{equation}
 where $\Gamma$ is the partial width of the decay channel.

 The measured \CP asymmetry  \calA an be written as   $\calA  = \ACP + \AFB + \AEP$,   where  \ACP\ is the direct \CP asymmetry contribution, \AFB is a forward/backward  asymmetry contribution in $c \bar{c}$ production from $\gamma-Z^0$ interference  and 
  higher order QED processes, and \AEP is the asymmetry contribution induced by the detector in tracking, particle identification, and in material interactions.  \AFB  asymmetry   is an odd function of  the cosine of the polar angle of the  \D(s)pm meson momentum in  the \epem\ center of mass (CM) system, $\cos \theta^*_D$.    \ACP and \AFB are both  measured while  data have been corrected  for \AEP\  with  a control sample.  The data-driven method used to correct for \AEP\ is  described in Ref.\cite{FB}. 
  
  A simultaneous binned  ML fit  to the $D^+_{(s)}$ and  $D^-_{(s)}$  invariant mass distributions is performed in 10 equally spaced bins of  $\cos \theta^*_D$ with bin 0 at [-1.0, -0.8].   Since \ACP is independent of the kinematic variable  $\cos \theta^*_D$, the asymmetry ${\cal A}(+| \cos \theta^*_D|$ ) measured in a positive  $\cos  \theta^*_D$ bin and the asymmetry  
  ${\cal A}(-| \cos \theta^*_D|$ ) measured in  its  symmetric (negative) counterpart  $\cos \theta^*_D$ bin give the same contribution to \ACP.  On the other hand since \AFB is an odd function of
   $\cos \theta^*_D$ , the contribution to \AFB from symmetric $\cos \theta^*_D$ bins have the same magnitude and opposte sign.  So \ACP and \AFB as a function of $\cos  \theta^*_D$  
  can be written in the form:

\begin{eqnarray}
\AFB(|\cos^* \theta_D|) &=&  \frac{{\cal A}(+|\cos \theta^*_D|)  -   {\cal A}(-|\cos \theta^*_D|)  }{2} \nonumber \\
\ACP(|\cos^* \theta_D|) &=&  \frac{{\cal A}(+|\cos \theta^*_D|)  +   {\cal A}(-|\cos \theta^*_D|)  }{2} \,,
\end{eqnarray}

\begin{figure}[htb]
\begin{center}
\includegraphics[scale=0.45] {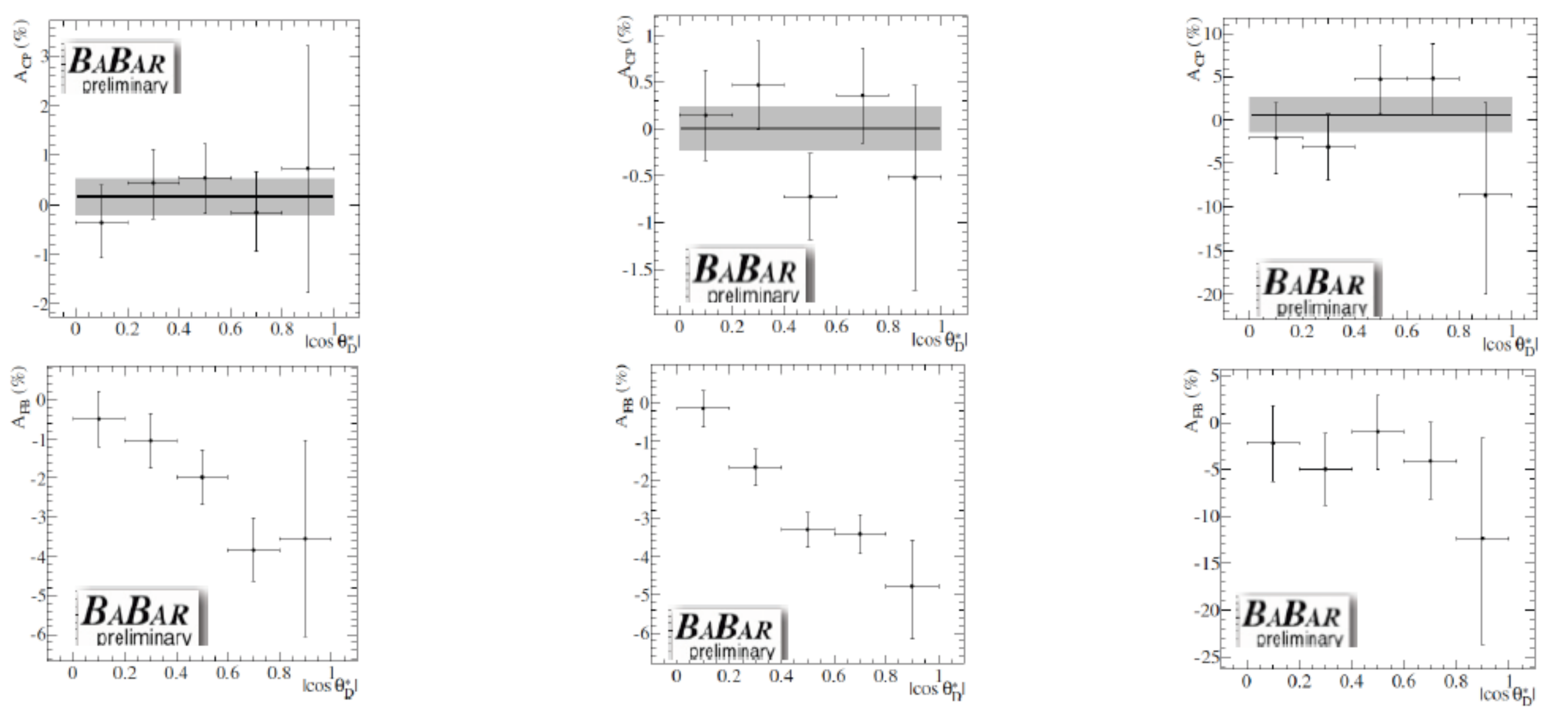} \\  
\caption{  \ACP (top) and \AFB (bottom) asymmetries for $\Dpm \ra \KS K^{\pm}$  (left), $\Dspm \ra \KS K^{\pm}$ (center), and $\Dspm \ra \KS \pi^{\pm}$  (right) as a function of $|\cos^* \theta_D|$ in the data sample. 
The solid line represents the central value of \ACP and the gray region is the $\pm \sigma$ interval, both from a $\chi^2$ minimization assuming no dependence of \ACP on $|\cos^* \theta_D|$.
}
 \label{fig:DCP}
\end{center}
\end{figure}

The values of \ACP and \AFB\  asymmetries are shown in Fig.~\ref{fig:DCP}.

For each decay mode Table~\ref{tab:tabella}  shows the  \ACP value from  the fit,  the   bias corrected \ACP value and in the last raw the \ACP value after subtracting the expcted 
\ACP contribution due to \Kz-\Kzb mixing.  These results are consistent with zero and with the \SM predictions  within 1 $\sigma$.

\begin{table*}[t]
\footnotesize
\caption{ $A_{\CP}$ measurements.
Uncertainties, where reported, are statistical the first and systematic the second (\babar\ Preliminary).}
\begin{center}
\begin{tabular}{|l|c|c|c|}\hline
  & \Dtoksk & \Dstoksk & \Dstokspi \\
\hline\hline
$A_{\CP}$ value from the fit & $(+0.16 \pm 0.36)\%$ & $(0.00 \pm 0.23)\%$ & $(+0.6 \pm 2.0)\%$ \\
\hline\hline
\multicolumn{4}{|l|}{Bias Corrections for:}  \\
 Toy  MC experiments & $+0.013\%$ & $-0.01\%$ & $-$ \\
PID selectors  & $-0.05\%$ & $-0.05\%$ & $-0.05\%$ \\
\KSKL\ interference   & $+0.015\%$ & $+0.014\%$ & $-0.008\%$ \\
\hline\hline
$A_{\CP}$ final value & $(+0.13 \pm 0.36 \pm 0.25)\%$ & $(-0.05 \pm 0.23 \pm 0.24)\%$ & $(+0.6 \pm 2.0 \pm 0.3)\%$\\
\hline\hline
$A_{\CP}$ contribution & $(-0.332 \pm 0.006)\%$ & $(-0.332 \pm 0.006)\%$ & $(+0.332 \pm 0.006)\%$ \\
from \KzKzb mixing &&&\\
\hline\hline
$A_{\CP}$ final value (charm only) & $(+0.46 \pm 0.36 \pm 0.25)\%$ & $(+0.28 \pm 0.23 \pm 0.24)\%$ & $(+0.3 \pm 2.0 \pm 0.3)\%$\\
\hline
\end{tabular}
\label{tab:tabella}
\end{center}
\end{table*}

 \section {CP Violation  in the Decays \DpmtoKKpipm }
Searches for direct \CP violation in the  SCS decays \DpmtoKKpipm have been recently performed by \babar, using a dataset 
of   $476 fb^{-1}$~\cite{Puro}. This  sample  contains enough 3-body  SCS decays to probe \CP at the level of  \SM predictions.  The decay \DptoKKpip~\cite{CC} is dominated by 
quasi-two body  decays  with resonant intermediate states, giving possibility to study direct \CPV in a particular resonance or in different regions 
of the Dalitz plot (\DP).   In previous analyses of these 3-body decay modes performed by  CLEO-c~\cite{CLEO-c} and LHCb~\cite{KKpi} Collaborations no evidence for   \CPV has been found   in agreement with \SM prediction.

Signal reconstruction efficiency is determined with a sample of Monte Carlo (MC)  simulated events from the distribution of reconstructed  events 
as a function of the  \CM polar angle of the D meson ($\cos \theta_{CM}$) and  of the $m^2(K^- \pi^+)$ vs $m^2(K^+K^-)$ \DP.  The ratio of efficiency-corrected signal yields,  
$R = \frac{  N_{D^+}/\epsilon_{D^+}    }{   N_{D^-}/\epsilon_{D^-}   } = 1.020 \pm 0.006$ is used to allow for asymmetries in the MC event production due to physics or
detector induced effects. 

Time-integrated \CP asymmetry (charge asymmetry)  is defined in a given bin  as:
\begin{equation}
{\calA} \equiv  \frac{N_{D^+}/\epsilon_{D^+ }   -   N_{D^-}/\epsilon_{D^- } }      {N_{D^+}/\epsilon_{D^+ }   +   N_{D^-}/\epsilon_{D^- } }
\end{equation}

\begin{figure}[htb]
\begin{center}
\includegraphics[scale=0.40] {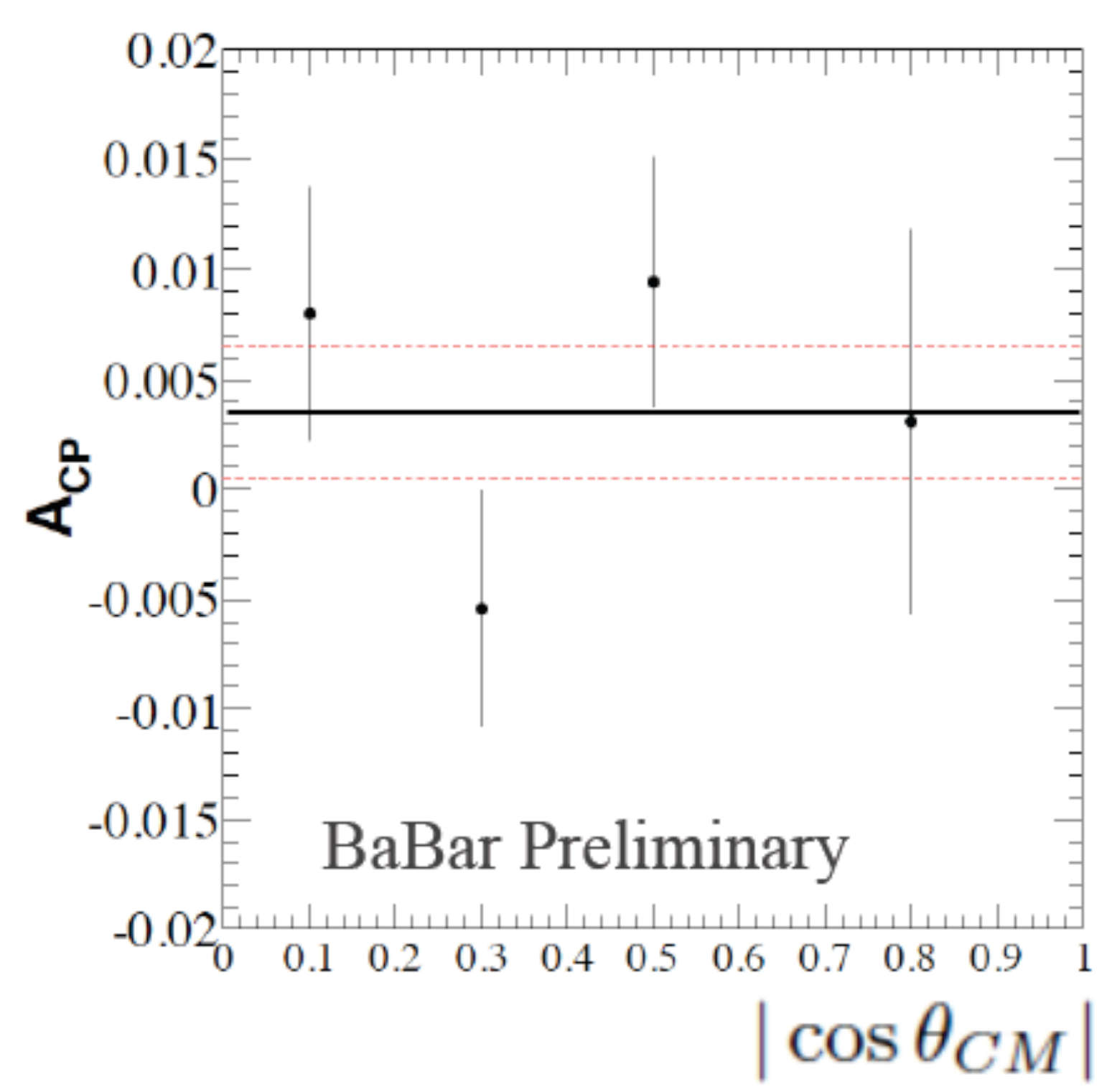} \\  
\caption{  Charge asymmetry \ACP as a function of  $|\cos \theta_{CM}| $ in data. 
The solid line represents the central value of \ACP and the dashed lines represent  the $\pm \sigma$ interval, both determined  from a $\chi^2$ minimization assuming no 
dependence on $|\cos \theta_{CM}|$.
}
 \label{fig:ChAs}
\end{center}
\end{figure}

 Selection efficiencies  are corrected to account for 
 differences between data and MC simulated events in the reconstruction asymmetry of  charged pion tracks and in the production model of charm mesons. The charge 
 asymmetry  \calA  contains contributions from both  the forward/backward asymmetry \AFB  and the direct \CP asymmetry \ACP.  To remove the  contribution \AFB,   the 
 charge asymmetry is averaged over four symmetric bins in $\cos \theta_{CM} $. The averaged  values  of  \ACP in the four bins  are shown Fig.~\ref{fig:ChAs}. The central 
 value $\ACP = (0.35 \pm 0.30 \pm 0.15)  $ \% is obtained with a $\chi^2$ minimization. The probability that the asymmetries are null in all the four bins is 21\%.

A model-independent technique to search for \CP violation in \DP is to compare \CP asymmetries in different regions of the \DP.  The results  of \ACP  asymmetry  measured in four 
 regions of \DP are given in  Table~\ref{tab:tab1}.  Measured \CP asymmetries are consistent with zero.  
 
 We also measure the normalized residuals  $\Delta$ of efficiency-corrected 
 and background subtracted \DP for \Dp and \Dm for equally populated  bins. $\Delta$ is defined as

\begin{equation}
\Delta \equiv \frac{ n(\Dp) - R\, n(\Dm)}  {\sqrt{\sigma^2(\Dp) + R^2 \sigma^2(\Dm)}}\,,
\end{equation}
where n is the yield in a bin in the \DP and $\sigma$ its uncertainty. 

$\Delta$ distribution  is fitted to a Gaussian function. For 100 bins we obtain a Gaussian residual mean of $0.08 \pm 0.15$ and a width of  $1.11 \pm 0.15$. 
The probability that the two \DP's are  consistent with no \CP asymmetry is 72\%.

\begin{table*}[]
\caption{ Yields, efficiencies,, and \CP asymmetry in  four different regions of the DP.  First uncertainty in the \CP asymmetry is statistical, the second  is systematic (\babar\ Preliminary).}
\begin{center}
\includegraphics[scale=0.34]{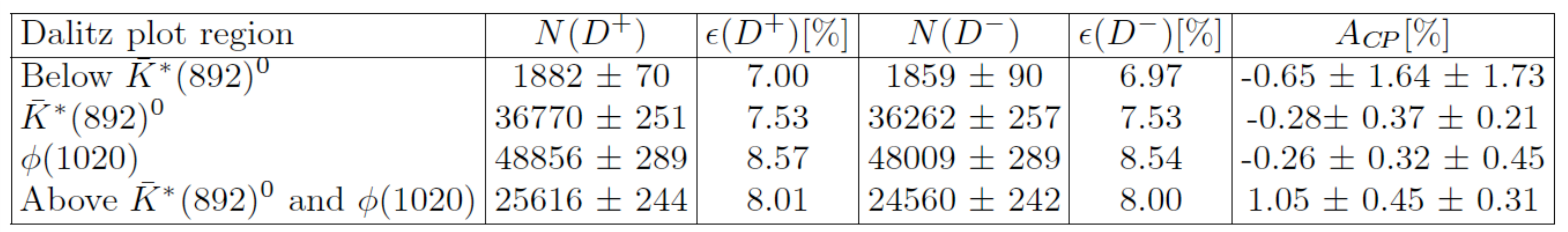} \\  
\label{tab:tab1}
\end{center}
\end{table*}

Angular moments of the cosine of the helicity angle  $\theta_H$  of the D decay products reflect the spin and mass  of intermediate resonant and non resonant states~\cite{babar1}.
The helicity angle $\theta_H$ in the decay $D\ra (r \ra AB)C $ is defined  as the angle between the momenta of B  and parent D in the AB  rest frame.
 We can search for \CP in the \DP  in a model-independent way by comparing  the angular moments between \Dp and \Dm~\cite{babar2}.  Angular moments  
 of order l are defined  as the efficiency-corrected and background-subtracted two-body invariant mass distributions ($m(K^+K^)-$, $m(K^- \pi^+$)) weighted by spherical 
 harmonic moments $ w^{(l)} = Y^0_l (\cos \theta_H)$.  Weights in two-body invariant mass intervals are defined as:
 
 \begin{equation}
 W^{(l)}_i  \equiv   \frac{\left(\sum_j w^{(l)S}_{ij} - \sum_k  w^{(l)B}_{ik}\right)}{<\epsilon_i>}\,,
 \end{equation} 
 where $i$ is bin index, and $j$, $k$ event indices. $S$ and $B$ refer to signal and background, and  $<\epsilon_i>$ is the average efficiency in bin $i$.
 
 Normalized moment residuals $X_l $ for \Dp and \Dm are calculated for l from 0 to 7:
 
 \begin{equation}
 X_l = \frac{\left(W^{(l)}_i (\Dp) -  R W^{(l)}_i (\Dm) \right)  }   { \sqrt{ \sigma_1^{ (l)^2   } (\Dp) + R^2 \sigma_1^{ (l)^2   } (\Dm)    }  }
 \end{equation}

The   $\chi^2$  is calculated over all the mass bins in $K^+K^-$ and $K^-\pi^+$ moments with

\begin{equation}
\chi^2 = \sum_i  \sum_{l_1} \sum_{l_2}  X_i^{(l_1)} \rho_i^{(l_1l_2)}   X_i^{(l_2)}\,,
\end{equation}
 where      $\rho_i^{(l_1l_2)}  $     is the correlation coefficient between $X_i^{(l_1)}$ and $X_i^{(l_2)}$.

With a number  of  degrees of freedom NDF equal   to 287  the $\chi^2/NDF$ in the $K^+K^-$ and $K^-\pi^+$ moments is  1.10 and 1.09, 
consistent with no \CPV at 11\% and 13 \%, respectively.

\babar\  also searched for \CPV in  a model-dependent \DP analysis of the \DptoKKpip decay~\cite{CC}.  The \DP amplitude \calA in the isobar model is written as a set  of two-body intermediate states r :
${\cal A} = \sum_r  {\cal M}_r   e^{i\phi_r} F_r $, 
where ${\calM}_r$ and $\phi_r$ are real and $F_r = F_r(m(K^+K^-), m(K^-\pi^+))$ are dynamical functions describing the intermediate states.
In case of  amplitudes  with  small contributions the complex  coefficient has been parameterized in a Cartesian form: $x_r = {\cal M}_r \cos \phi_r$ and $y_r = {\calM }_r \sin \phi_r$.
The  $K^*(892)^0$ has been chosen as the reference amplitude. Assuming no \CPV the relative fractions  of resonances  and a constant non resonant amplitude over the entire \DP contributing to the decay have been determined with an unbinned ML fit.

To allow for possible \CPV  in the decay,  the resonances of the \Dp (\Dm) decays  contributing with a fit fraction of at least 1\%  have been parameterized with different amplitudes and phases in their decay amplitudes. 
A simultaneous fit  to the \Dp and \Dm samples have been performed, parameterizing each resonance with four parameters, ${\calM}_r$, $\phi_r$, $r_{\CP}$, and $ \Delta\phi_{\CP}$.  The \CPV parameters are $r_{\CP} = \frac{ |{\calM}_r|^2 - |{\bar{\calM}}_r|^2 }   { |{\calM}_r|^2 + |{\bar{\calM}}_r|^2 }$ and   $ \Delta\phi_{\CP} = \phi_r - \bar{\phi}_r$.
The Cartesian form   of  the \CP  violating parameters  are $\Delta x_r$ and $\Delta y_r$ with  $x_r(\Dpm)  = x_r \pm \Delta x_r/2 $ and $ y_r(\Dpm) = y_r \pm \Delta y_r/2 $. Fit results are shown in Table~\ref{tab:tab2}.

All \CPV  parameters from \DP fit are consistent with zero and with \SM expectations.

\begin{table*}[]
\caption{ \CPV parameters from the \DP fit.  First uncertainties are statistical, the second systematic (\babar\ Preliminary).}
\begin{center}
\includegraphics[scale=0.25] {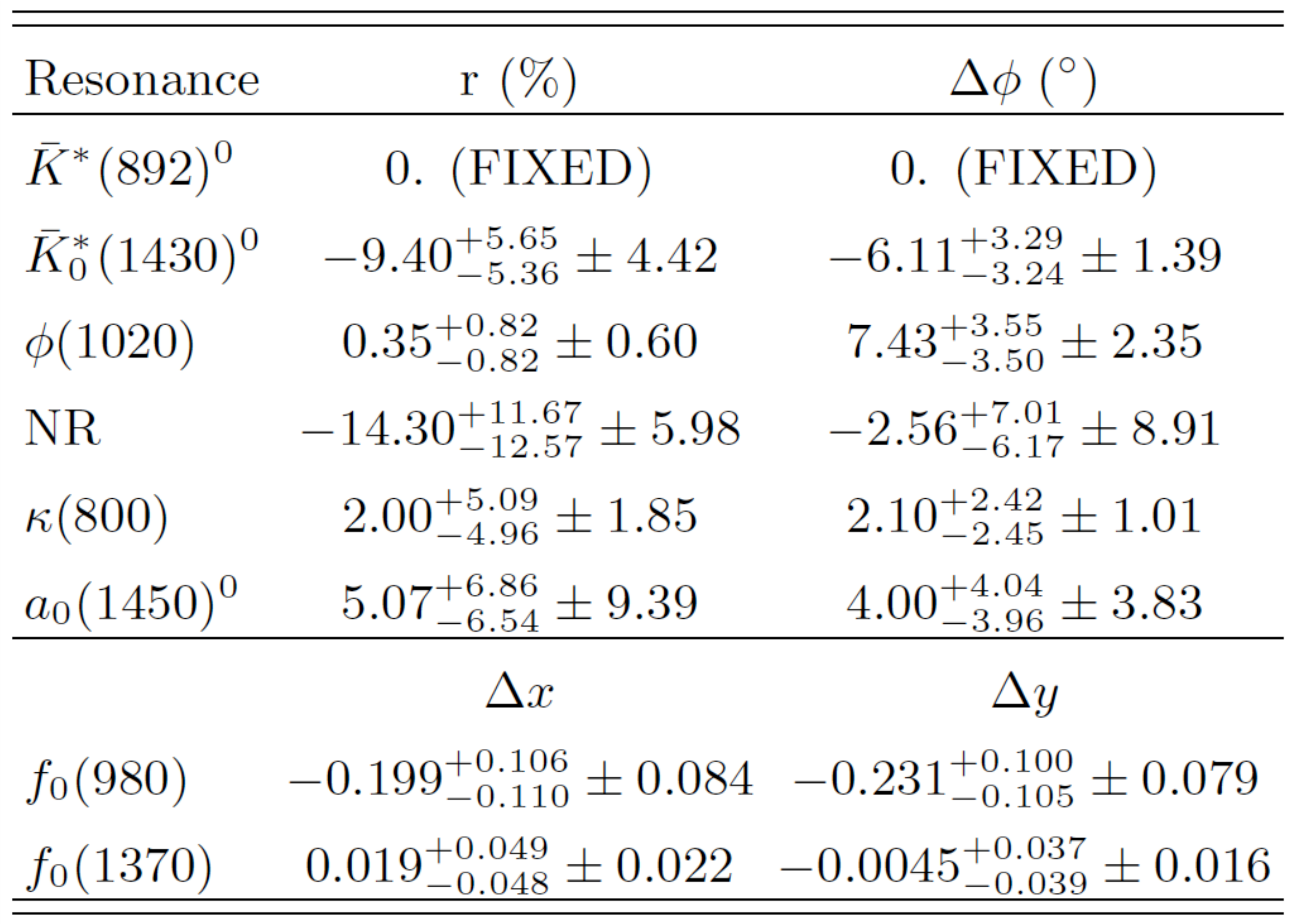} \\  
\label{tab:tab2}
\end{center}
\end{table*}

 \section {Conclusions} 
I have presented recent  improved \babar\ measurement of mixing and a search of \CP violation in two-body \Dz decays, a search for  direct \CP violation  in  \Dtoksk 
and   \Dstoksk, \kspipm, and  searches for \CP violation in the decays \DpmtoKKpipm  both using model-independent and model-dependent analysis techniques.
All results in these analyses are well described  within the \SM and no effect related to \NP has been found. The measured mixing parameter $y_{\CP} = [0.72 \pm  0.18 (stat) \pm 0.12 (syst)]$\% excludes  the  no-mixing null hypothesis   with a significance of  3.3$\sigma$.

\section {Acknowledgments}
I wish to warmly thank the organizers of QUARKS-2012 for the excellent program and for the kind hospitality in Yaroslavl. I would like to thank also G. Casarosa for useful comments.

\end{document}